\def\arcdeg{$^{\circ}$}
\def\arcmin{'}

\documentstyle[12pt,epsfig]{article}
\renewcommand{\section}[1]{\vspace{6pt} \noindent\mbox{#1} \newline \noindent}
\renewcommand{\subsection}[1]{\vspace{6pt} \noindent\mbox{\underline{#1}} 
\newline \noindent}
\renewcommand{\subsubsection}[1]{\vspace{6pt} \noindent\mbox{\underline{#1}}
\noindent}
\newfont{\sansb}{cmssbx10}
\newfont{\sans}{cmss10}

\begin{document}
{\small OG 4.1.2 \vspace{-24pt}\\}     
{\center \LARGE OBSERVATION OF SPECTRUM OF TEV GAMMA RAYS UP TO 60 TEV 
FROM THE CRAB AT THE LARGE ZENITH ANGLES
\vspace{6pt}\\}
{\center
K.Sakurazawa$^1$,~T.Tanimori$^1$,~S.A.Dazeley$^2$,~P.G.Edwards$^3$,~S.Hara$^1$,
~T.Hara$^4$,~Y.Hayami$^1$, \\
S.Kamei$^1$,~T.Kifune$^5$,~R.Kita$^6$,~T.Konishi$^7$,~A.Masaike$^8$,
~Y.Matsubara$^9$,~Y.Matsuoka$^9$, \\
Y.Mizumoto$^{10}$,~M.Mori$^{11}$,~H.Muraishi$^6$,~Y.Muraki$^9$,~T.Naito$^{12}$,
~K.Nishijima$^{13}$,~S.Ogio$^1$, \\
J.R.Patterson$^2$,~M.D.Roberts$^2$,~G.P.Rowell$^2$,~T.Sako$^9$,
~R.Susukita$^{14}$,~A.Suzuki$^7$,~R.Suzuki$^1$, \\
T.Tamura$^{15}$,~G.J.Thornton$^2$,~S.Yanagita$^6$,~T.Yoshida$^6$,
~T.Yoshikoshi$^5$ \vspace{6pt} \\
}

{\it $^1$Department of Physics,Tokyo Institute of Technology,Tokyo 152,Japan\\
$^2$Department of Physics and Mathematical Physics, University of Adelaide, SouthAustralia 5005,Australia \\
$^3$Institute of Space and Astronomical Science, Sagamihara 229,Japan \\
$^4$Faculty of Commercial Science, Yamanashi Gakuin University,Kofu 400,Japan \\
$^5$Institute for Cosmic Ray Research,University of Tokyo,Tokyo 188,Japan \\
$^6$Faculty of Science,Ibaraki University,Mito 310,Japan \\
$^7$Department of Physics,Kobe University,Kobe 637,Japan \\
$^8$Department of Physics,Kyoto University,Kyoto 606,Japan \\
$^9$Solar-Terrestrial Environment Laboratory,Nagoya University,Nagoya 464,Japan \\
$^{10}$National Astronomical Observatory,Tokyo 181,Japan \\
$^{11}$Faculty of Education,Miyagi University of Education,Sendai 980,Japan \\
$^{12}$Department of Physics,Faculty of Science,University of Tokyo,Tokyo 113,Japan \\
$^{13}$Department of Physics,Tokai University,Hiratsuka 259,Japan \\
$^{14}$Institute of Physical and Chemical Research,Wako,Saitama 351,Japan \\
$^{15}$Faculty of Engineering,Kanagawa University,Yokohama 221,Japan
\vspace{-12pt} \\}
{\center ABSTRACT\\}
The CANGAROO experiment has observed gamma-ray above 7TeV from
the Crab pulsar/nebula at large zenith angle
in Woomera, South Australia.
We report the CANGAROO data taken in 1992, 1993 and 1995,
from which it appears that the energy spectrum extends at least up to 50 TeV.
The observed integral spectrum is
$(8.4\pm1.0)\times 10^{-13}(E/7\,{\rm TeV})^{-1.53 \pm 0.15}$\ 
cm$^{-2}$s$^{-1}$ between 7 TeV and 50 TeV.

In November 1996, the 3.8m mirror was recoated in Australia,
and its reflectivity was improved to be  about 90\%  as twice as before.
Due to this recoating, 
the threshold energy of $\sim$ 4 TeV for gamma rays 
has been attained in the observation of the Crab at large zenith angle.
Here we also  report the preliminary result  taken in 1996.

\setlength{\parindent}{1cm}
\section{INTRODUCTION}
The Crab pulsar/nebula has been observed by various energy regions.
In the energy region from radio to X-ray, it is well understood as
synchrotron emission from high energy electrons accelerated up to $\sim$100
TeV in the nebula.
The higher energy component above $\sim$1 GeV is considered to
be produced by the inverce Compton scattering between these very high
energy electrons and ambient photons in the nebula.

Previously following energy regions in the range of IC gamma-rays 
have been observed.
The 0.1--10GeV region has been observed by EGRET,
and 0.2--10TeV region, by some imaging air \v{C}erenkov telescopes(IACT).
In particular, there are only two dataset around 10 TeV:
one is our previous data above 7 TeV with $\sim$4$\sigma$ level uncertainly
and the other is the data up to 12 TeV reported by the Themistocle group.
The observation of gamma-rays above 10 TeV is the key to understand
the IC process in the nebula.

\section{OBSERVATION}
The observation was made with the 3.8m telescope of the CANGAROO
collaboration between Japan and Australia,
which is located at Woomera in South Australia (136\arcdeg47\arcmin E and
31\arcdeg06\arcmin S).
The high resolution imaging camera, set at the prime focus, 
consists of small square-shaped 
photomultiplier tubes of 10mm $\times$ 10mm size (Hamamatsu R2248).
The number of photomultipliers was  220 in 1993
and increased to 256 in 1995 with a perfect square alignment, 
and giving a total field of view of about 3\arcdeg.
The details of the camera and telescope have
already been described in Hara et al.\ (1992).
The reflectivity of the mirror has estimated about 45\% in 1994.
In November 1996, the mirror was recoated in Angro-Australian Observatory,
and  the reflectivity of the mirror incresed up to about 90\%.

The  Crab  was observed at zenith angles
of 53\arcdeg--56\arcdeg in 1992, 1993, 1995 and 1996.
In order to monitor the cosmic ray background contained in `ON-source'
data, `OFF-source' runs were also done as described 
in Tanimori et al.\ (1994).
The total observation times for on- and off-source runs were $3.51\times10^5$~s
and $2.92\times10^5$~s, respectively.

\section{ANALYSIS AND RESULT}
The imaging analysis procedure applied to the data
is similar to that described in Tanimori et al.\ (1994). 
Basically, we use the same parameters characterizing 
the elongated shape of the \v{C}erenkov
light image as those used by the Whipple group (Weekes et al.\ 1989):
``width'',``length'', ``distance'',
``conc'', and the orientation angle $\alpha$.
The resulting  event distributions in the combined 1992, 1993 and 1995 
observations are plotted in Fig.1a as a function of $\alpha$,
and the preliminary one in two nights observation in 1996 
are plotted in Fig.1b.
The number of background events in the $\alpha$ peak region
$(\alpha\le 15^{\circ})$
was estimated from the flat region of $\alpha$ distribution
(30\arcdeg--90\arcdeg) of the on-source plot. 
The statistical significance of the   
peak in Fig.1a achieved is 8$\sigma$.
Effects by the bright star ($\zeta$ Tau, visual magnitude 3.0), 
which is located at 1.1\arcdeg \ from
the Crab and within the field of view of the camera, were
investigated by the same procedure applied to  1992 
observation. 
We found no effect which would cause false $\alpha$ peaks.

\begin{figure}[h]
\begin{center}
\epsfig{file=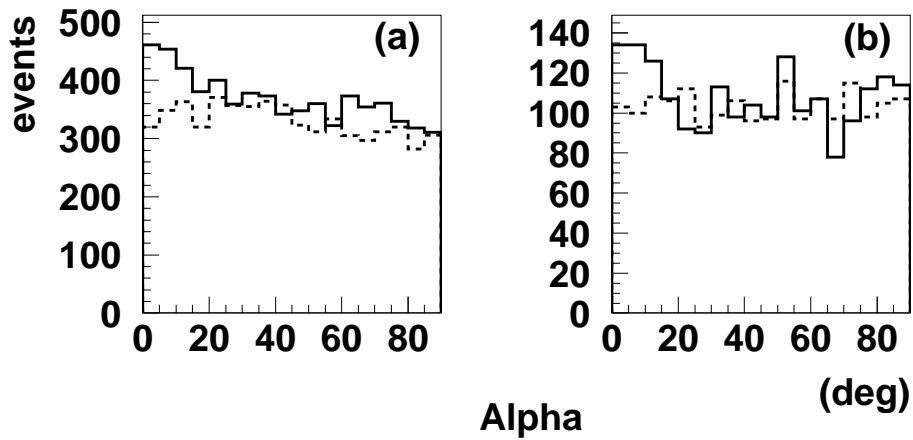,height=7.2cm}
\caption{$\alpha$ distribution. (a): until 1995, (b): in 1996.}
\end{center}
\end{figure}

The collecting  area and threshold energy of the telescope for gamma-ray
showers have been inferred from a Monte Carlo simulation as described in 
Tanimori et al.\ (1994).
The integral fluxes of gamma-rays observed
are $(8.0\pm1.1)\times10^{-13}$ cm$^{-2}$s$^{-1}$ above $\sim$7 TeV until 1995,
and $(2.2\pm0.6)\times10^{-12}$. cm$^{-2}$s$^{-1}$ above $\sim$4 TeV in 1996.
The threshold energy is defined as the energy
of the maximum differential flux of gamma-ray events
expected to be detected by the Monte Carlo simulation.

In order to obtain the integral energy spectrum,
$\alpha$ plots were made by varying the minimum number of 
detected photons,
required in the event analysis.
The threshold energy corresponding to each $\alpha$ plot
was estimated from the simulation.

The resultant integral flux spectrum 
between 7 TeV and 50 TeV
is plotted in Fig.2,  and can be written in the form of
$(8.4\pm 1.0)\times 10^{-13}(E/{7\,\rm TeV})^{-1.53 \pm 0.15}$ 
cm$^{-2}$s$^{-1}$.
The absolute flux above a few TeV  matches well
both of the results recently revised  
by the Themistocle Group (Djannati-Atai et al.\ 1995) 
and the Whipple Group (Weekes et al. 1996).
The integral index of $-1.55$ is also consistent with 
that of the Themistocle group of $-1.4$ obtained between 2 TeV and 12 TeV
and that of the Whipple group of $\sim -1.5$ from 500 GeV 
and up to $\sim$10 TeV within 1 $\sigma$ error.
The uncertainties of the energy of the gamma-ray is  about 25\%,
which is due to mainly the errors of the estimations 
of the  number of detected photo-electrons and 
the reflectivity of the reflector.

\begin{figure}[h]
\begin{center}
\epsfig{file=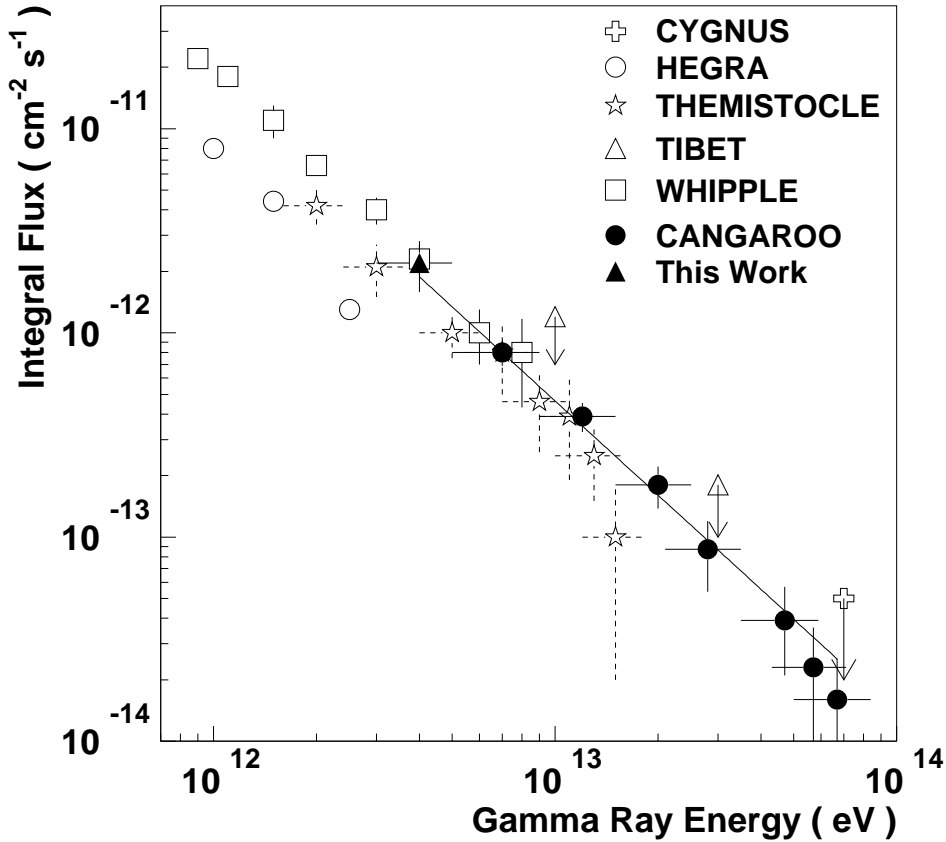,height=13.5cm}
\caption{Integral Energy Spetrum of Crab pulsar/nebula }
\end{center}
\end{figure}

In addition, the integral spectrum of the background,
cosmic rays, was calculated from the data, which were cut only ``distance'',
to be  $(9.4\pm 0.3)\times 10^{-8}(E/{10\,\rm TeV})^{-1.72 \pm 0.04}$ 
cm$^{-2}$s$^{-1}$ in the energy range of 13TeV to 110TeV,
where all of background events are considered to be proton showers.
Not only the integral index of cosmic rays 
can be reconstructed with good accuracy,
but also the absolute flux is fairly consistent  
to the recent results of balloon born
experiments within factor $\sim$3.
Here the same procedure of the simulation was used to calculate the
efficiencies of the trigger and analysis for proton showers.

\section{ACKNOWLEDGEMENTS}
This work is  supported by a Grant-in-Aid in Scientific Research
of the Japan Ministry of Education, Science, Sports  and Culture,
and also by the Australian Research Council and International Science
and Technology Program.
T.T. and T.K. acknowledge the support of the Sumitomo Foundation.
The receipt of JSPS Research Fellowships (P.G.E., T.N., M.D.R.,
G.J.T., T.Y. and K.Sakurazawa) is also acknowledged.

\section{REFERENCES}
\setlength{\parindent}{-5mm}
\begin{list}{}{\topsep 0pt \partopsep 0pt \itemsep 0pt \leftmargin 5mm
\parsep 0pt \itemindent -5mm}
\vspace{-15pt}
\item Alexandreas, D.~E., et al.\ 1991, ApJ, 383, 653
\item Amenomori, M., et al. 1992, Phys.Rev.Lett., 69, 2468
\item Atoyan, A.~M., \& Aharonian, A.~F. 1996, MNRAS, 278, 525
\item Cheung, W.M., \& cheng, K.S., 1994, ApJ, 90, 827 
\item Djannati-Atai,A.  et al. 1995, Proc.\ 24th Internat.\ Cosmic 
  Ray Conf.\ (Rome), 1, 315
\item De Jager, O.~C., \& Harding, A.K.\ 1992, ApJ, 396, 161
\item De Jager, O.~C., et al.\ 1996, ApJ, 457, 253 
\item Goret, P., et al.\ 1993, A\&A, 270, 401
\item Hara, T., et al.\ 1993 Nucl.\ Instr.\ and Meth., A332, 300
\item Hester, J.J., et al.\ 1995, ApJ, 448, 240
\item Ichimura, M., et al. 1993, Phys.Rev., D48, 1949
\item Konopelko, A., et al.\ 1996, Astroparticle Phys., 4, 199
\item Nolan, P.~L., et al.\ 1993, ApJ, 409, 697
\item Patterson, J.~R., \& Kifune, T.\ 1992, Australian and 
  New Zealand Physicist, 29, 58
\item Petry, D., et al.\ 1996, A\&A, 311, 13
\item Vacanti, G., et al.\ 1991, ApJ, 377, 467
\item Weekes, T.~C., et al.\ 1989, ApJ, 342, 379
\item Weekes, T.~C., et al.\ 1996, VERITAS proposal, Appendix A, 8
\end{list}

\end{document}